*Article*

# Examining and Comparing the Effectiveness of Virtual Reality Serious Games and LEGO Serious Play for Learning Scrum

Aldo Gordillo * , Daniel López-Fernández  and Jesús Mayor

Computer Science Department, Universidad Politécnica de Madrid, 28040 Madrid, Spain;
daniel.lopez@upm.es (D.L.-F.); jesus.mayor@upm.es (J.M.)
* Correspondence: a.gordillo@upm.es

**Abstract:** Significant research work has been undertaken related to the game-based learning approach over the last years. However, a closer look at this work reveals that further research is needed to examine some types of game-based learning approaches such as virtual reality serious games and LEGO Serious Play. This article examines and compares the effectiveness for learning Scrum and related agile practices of a serious game based on virtual reality and a learning activity based on the LEGO Serious Play methodology. The presented study used a quasi-experimental design with two groups, pre- and post-tests, and a perceptions questionnaire. The sample was composed of 59 software engineering students, 22 of which belonged to group A, while the other 37 were part of group B. The students in group A played the virtual reality serious game, whereas the students in group B conducted the LEGO Serious Play activity. The results show that both game-based learning approaches were effective for learning Scrum and related agile practices in terms of learning performance and motivation, but they also show that the students who played the virtual reality serious game outperformed their peers from the other group in terms of learning performance.

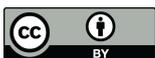

## 1. Introduction

A considerable effort has been and continues to be devoted to search for new alternatives to improve and complement traditional learning methodologies. In recent years, game-based learning has become one of the most promising learning approaches due to its potential to increase not only students' motivation but also students' learning. As evidenced by several recent literature reviews [1–7], a plethora of studies reported successful game-based learning experiences. This evidence allows the conclusion that, when game-based learning is suitably applied, it can lead to enhancements in both students' motivation and learning outcomes.

However, although a lot of research work has been conducted in the game-based learning field, a closer look at this work reveals that not all types of game-based learning approaches have been extensively investigated. For example, despite there being plenty of studies examining the educational use of videogames [1–7], much less research has been conducted to examine the use of educational videogames based on virtual reality or game-based learning activities based on LEGO Serious Play.

This article examines and compares the instructional effectiveness of a virtual reality serious game and a learning activity based on the LEGO Serious Play methodology. In particular, the study reported in this article examined and compared the use of the two approaches for learning the Scrum framework and some agile practices commonly adopted by Scrum teams. In this regard, it should be noted that, nowadays, Scrum is by far the most widely employed agile methodology in the industry [8] and is part of many degree programs, especially in software engineering. The research questions addressed by this article are as follows:

- RQ1: Are virtual reality serious games effective in terms of knowledge acquisition and motivation for learning Scrum?
- RQ2: Are activities based on LEGO Serious Play effective in terms of knowledge acquisition and motivation for learning Scrum?
- RQ3: Are virtual reality serious games more effective than activities based on LEGO Serious Play in terms of knowledge acquisition and motivation for learning Scrum?

The article is organized into six sections. The next section reviews the existing literature on virtual reality serious games, LEGO Serious Play, and Scrum training. Section 3 details the research methodology, including a description of the virtual reality serious game and the LEGO Serious Play activity that were employed. The results are presented and discussed in Sections 4 and 5, respectively. Finally, Section 6 outlines the conclusions of the article and suggests future works that could be undertaken.

## 2. Related Work

### 2.1. Serious Games Based on Virtual Reality

As evidenced by recent literature reviews, virtual reality is being widely used for educational purposes in many knowledge fields and industrial sectors because this technology contributes to improve knowledge acquisition and skills development, engages and motivates learners, and enhances the whole learning experience [9–11].

Some works have reported the use of virtual reality serious games by practitioners in professional contexts, including games to simulate medical operations [12], operate control panels in power plants [13], and conduct psychological exposure therapies [14]. Furthermore, some works have reported the use of this kind of game in educational settings, including games for learning languages and culture [15], physics [16], engineering [17], and topics related to computer science such as computer programming [18]. In this regard, it should be pointed out that the authors of the present contribution recently presented ScrumVR [19–21], which is the virtual reality serious game examined in this article.

In addition to ScrumVR, other virtual reality serious games aimed to teach the Scrum framework have been reported in the literature [22–24]. In the game presented by Caserman et al. [22], the player controls a Scrum Master that must guide their team throughout a Sprint. Radhakrishnan and Koumaditis [23] proposed the creation of a multiplayer virtual reality environment for simulating Sprints. However, a full implementation or validation of this system was not reported in that work. In a more recent work, Visescu et al. [24] presented ScrumSim, a multiplayer simulation in which players, with the assistance of two avatars who play the roles of the Scrum Master and the Product Owner, interact with a virtual reality environment, perform tasks, and receive feedback on their performance. Although ScrumSim has been used in real-world settings, a formal evaluation has not yet been reported.

### 2.2. LEGO Serious Play

LEGO Serious Play [25] is a learning methodology initially designed to teach professionals in a playful and active way soft skills such as leadership, communication, and conflict resolution. However, in the last decades this methodology has evolved substantially [26] and it is being applied in higher education across many knowledge fields. Indeed, LEGO Serious Play activities have been used in fields as diverse as marketing [27], arts [28], industrial engineering [29], and systems engineering [30]. These contributions indicate that LEGO Serious Play can be used not only to learn soft skills, but also to learn specific competences in many fields of knowledge, including software engineering.

Several LEGO Serious Play activities have been designed to teach concepts related to software engineering. A very prolific author in this area is Kurkovsky [31–35], whose works present activities to teach requirements engineering [31,32], complex systems dependability [31], component integration, and software interface design [33], Test-Driven Development (TDD) [34], and Scrum [35]. The conclusion of these contributions is that this learning methodology is useful to learn specific competences about software engineering, develop soft skills, and improve students' motivation. In the same vein, some authors of the present contribution reported in [36] another LEGO Serious play which proved to be highly appealing and motivating from the students' perspective and effective to learn about software life-cycle models and software development activities.

Another remarkable LEGO Serious play activity in the software engineering field is LEGO City, which was conceived by Krivitsky [37] and serves to teach the Scrum framework. In his book [37], this author presents guidelines to conduct the activity as well as empirical experiences reporting positive outcomes. Beyond the experiences described by Krivitsky himself, this activity has been adapted and replicated by other researchers in professionals and higher education contexts. For example, ref. [38] used the LEGO City activity with professionals of Croatian IT companies, ref. [39] with students of a computer science degree delivered by the University of Pernambuco (Brazil), and ref. [40] with students of a software project management master degree delivered by the Aalto University (Finland). In every case, the authors found out that the LEGO City activity was very well appreciated by the participants and highly effective to learn about Scrum. Another interesting experience is the one reported by [41], who adapted the LEGO City activity to be performed remotely using online whiteboards and real-time communication tools. Similarly, ref. [42] adapted a LEGO Serious Play activity aimed at teaching Scrum to be used with Minetest, an open-source variant of the Minecraft game.

### 2.3. Scrum Training

As described in the previous subsections, virtual reality serious games [19–24] and LEGO Serious Play activities [35,37–42] have been used to deliver Scrum training. Moreover, other noteworthy alternatives have been used for this purpose, including serious video games not based on virtual reality [43–45], physical games [46–48], agile project management tools [49], and classroom exercises [50].

Scrum-X [43] is a game developed in Microsoft Excel 365 with VBA programming designed to teach Scrum to graduates and professionals in which players can manage software projects. Another serious video game for learning Scrum is Virtual Scrum [45]. In this game, players can explore and interact with a virtual world that simulates the room of a Scrum team. In [44], an online Scrum simulation conducted by using the multiplayer video game "Don't Starve Together", Trello, and Discord is presented. In addition to video games, some physical games aimed at teaching Scrum have been designed such as the card games PlayScrum [46] and "Don't Break the Build" [47]. Another physical game to teach Scrum was proposed by [48]. In this game, players organized in groups must build paper boats, hats, and planes following the Scrum framework.

Another approach that has been adopted for Scrum training is the planning and monitoring of software projects through agile project management tools such as Taiga [49]. Lastly, it is also worth mentioning that some classroom exercises have been proposed for introducing Scrum. For example, ref. [50] proposed "Play Ball", an exercise targeted to undergraduates that only requires 20–30 hand-size balls per team and that aims to introduce students to basic Scrum concepts and allow them to practice self-organization.

## 3. Research Methodology

The study presented in this article used a quasi-experimental design with two groups (A and B), pre- and post-tests, and a perceptions questionnaire. The goal of the study was to empirically examine and compare the effectiveness in terms of knowledge acquisition and motivation of two game-based interventions for learning Scrum and related agile practices: one in which participants played a virtual reality serious game and another based on LEGO Serious Play in which participants built a part of a city by using LEGO blocks. The students in group A played the virtual reality serious game, whereas the students in group B conducted the LEGO Serious Play activity.

### 3.1. Sample

This study involved the participation of 59 students in total. Group A was comprised of 22 students with a median age of 22.3 (SD = 2.3): 15 males, 5 females, and 2 students who preferred not to indicate their gender. Group B was comprised of 37 students with a median age of 22.5 (SD = 4.8): 32 males and 5 females. All students who participated in this study were enrolled in a project management course at Universidad Politécnica de Madrid. This project management course is worth 6 ECTS (European Credit Transfer System) credits (so it requires 150–180 h of student work) and all students in their fourth year of the Bachelor's Degree in software engineering are required to take it. The course covers the fundamentals and common practices of project management and introduces software project management using traditional and agile approaches.

### 3.2. Procedure

First, all participating students gave informed consent to participate in the study. Then, these students were split into two groups. Random assignment was not possible because some of the students had already played the serious game the previous academic year in a different course. Therefore, the students who had never played the serious game were assigned to group A and the remaining students were assigned to group B. Once the groups were established, all the participating students were given 10 min for completing a pre-test. Once this test was completed, students in group A played the virtual reality serious game over 40–45 min after spending 15 min attending to the teachers' instructions and setting up their devices; whereas students in group B conducted the LEGO Serious Play activity, which lasted around one hour. Then, all students were given 10 min for completing the post-test. Finally, students completed the perceptions questionnaire. The whole intervention, including the tests and the questionnaire, lasted around one hour and a half in both groups. All participating students completed both the pre- and post-test, as well as the perceptions questionnaire.

### 3.3. Methods and Instruments

The pre-test consisted of 10 theoretical questions encompassing the Scrum framework and related agile practices, including user stories, MoSCoW, story points, and Planning Poker. The questions were aligned with the official Scrum Guide 2020 [51] and the resources provided by the Agile Alliance [52]. Each question had four options, only one of which was correct. Correct answers were worth +1 point each, whereas incorrect answers were worth −0.33 points each. Thus, the pre-test was scored from 0 to 10. The post-test was composed by the same 10 questions included in the pre-test and was scored in the same way as this test. In this regard, it should be clarified that, with the aim of preventing students from memorizing answers, correct answers and additional feedback were only revealed to the students after the time to complete the post-test had expired. Moreover, in order to discourage copying and other unexpected misconduct, the students' final course grade remained unaffected by either their post-test scores or their pre-test scores.

The questionnaire distributed to all participants at the end of the intervention to collect their perceptions was developed ad hoc for this study. This questionnaire was comprised of two demographic questions (age and gender), a set of statements to be rated on a Likert scale ranging from one (strongly disagree) to five (strongly agree), and an open-ended question that allowed respondents to provide comments. The Likert items of the questionnaire are included in Section 4 together with the results. The questionnaire distributed to both groups was identical, except for item 18, which was exclusively included in the questionnaire given to the students in group A. This particular item aimed to inquire whether these students had experienced any dizziness while using the virtual reality serious game.

### 3.4. Materials: Virtual Reality Serious Game

The virtual reality serious game examined in this article is called ScrumVR and aims to instruct the player in the Scrum framework and some related agile practices in an immersive and effective way. ScrumVR is a first-person game in which the main character is a developer who starts to work in a software company that joins a Scrum team. During the game, the player receives explanations, interacts with other members of the Scrum team, engages actively in the Scrum events, contributes to the development of Scrum artifacts, and employs diverse agile techniques, including MoSCoW and Planning Poker. Screenshots of certain key events taking place throughout the game are shown in Figure 1.

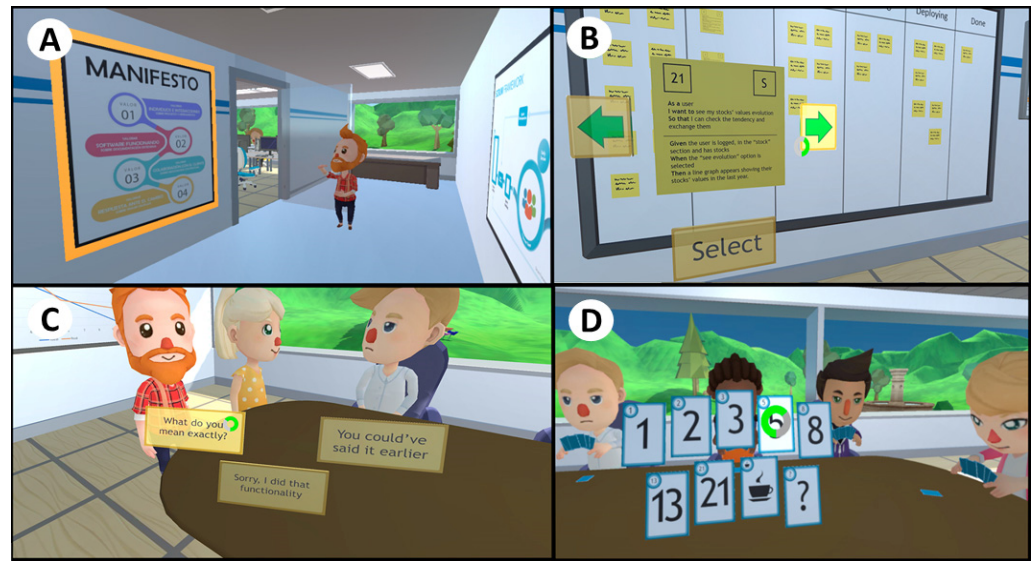

**Figure 1.** (**A**) The Scrum Master explains the foundation of agile methodologies. (**B**) The player chooses a user story. (**C**) The player talks with the Product Owner. (**D**) The player selects a card during a Planning Poker.

From a didactic standpoint, the game is narrated linearly, and it is divided into four chapters of approximately the same duration. Completing the four chapters requires the player to spend around 40–45 min. Each chapter is aimed at achieving certain learning objectives in such a way that the player learns how the Scrum framework works. Table 1 presents a summary of the main topics covered by each chapter of ScrumVR.

**Table 1.** Main topics covered by ScrumVR.

| Chapter | Topics |
| --- | --- |
| 1: Introduction | Agile and Scrum foundation.<br>The role of the Scrum Master.<br>How to operate a Kanban board. |
| 2: Daily Scrum | Performance of a Daily Meeting in Scrum.<br>How to select user stories based on their priority and size estimation.<br>How to interpret a burndown chart. |
| 3: Sprint Review and Sprint Retrospective | The role of the Product Owner.<br>Performance of a Sprint Review.<br>Performance of a Sprint Retrospective.<br>How to analyze the Sprint performance using a burndown chart. |
| 4: Sprint Planning | Performance of a Sprint Planning.<br>How to define user stories.<br>How to prioritize user stories using MoSCoW.<br>How to estimate the size of user stories using Planning Poker and story points. |

Regarding its technical characteristics, ScrumVR was developed using the Unity game engine and was conceived to be played by using Android devices equipped with a simple cardboard. One of the most remarkable points in the development of this educational application is the selected locomotion method. Since ScrumVR was designed for mobile devices, where positional tracking is not available, it was decided to use the on-rails guided locomotion method [53] in order to fully guide the player's movement and minimize the cybersickness feeling. Therefore, the players can look in all directions by rotating the head but do not have the freedom to move wherever they want, thus avoiding distractions and, consequently, facilitating the achievement of the established learning objectives. Additionally, positional sound [54] was used to guide the players to look in the most convenient direction throughout the experience. Another technical feature that was used for guiding the attention of the players was the use of illumination [55].

Another key technical feature of ScrumVR is the use of diegetic menus to allow the player to select an element by looking at it for a certain period of time. Whenever the player can perform an action in the game, a reticle that can be moved by rotating the head is displayed on the screen. When the reticle is placed over a selectable element, it starts to visually indicate the waiting time required to select that element. The diegetic menus used by ScrumVR allow the player to make decisions (see Figure 1C) and engage in agile techniques with other characters by playing mini-games (see Figure 1B,D). These mini-games were developed using a finite state machine model where all possible answers have an associated narrative feedback. Thereby, the learning experience is enriched with feedback generated according to the player's actions. More technical details about ScrumVR can be found at [19,20].

### 3.5. Materials: LEGO Serious Play Activity

The LEGO Serious Play activity assessed in this article aims to teach in an active and engaging way the Scrum framework and some related agile practices, including MoSCoW and Planning Poker. This activity was designed based on the LEGO City activity, which was initially conceived by Krivitsky [37] as explained in the related work section.

In the LEGO Serious Play activity, students work in teams of 4–6 members and each team assumes the role of a Scrum team belonging to a certain organization and working on a project whose goal is to build a part of a city (e.g., a neighborhood, a zoo, or a motor vehicle fleet). One student of each team assumes the role of the Product Owner, another student of the team assumes the role of the Scrum Master, and the remaining students of the team assume the role of the developers.

Each part of the city consists of a set of LEGO constructions. Therefore, each Scrum team has to build a part of a city by using the Scrum framework as methodology and LEGO pieces as materials. At the beginning of the activity, the Product Owner of each team is provided with a set of user story cards. Each of these cards includes a user story that describes a LEGO construction following the classic role–feature–benefit pattern, as well as two empty boxes in order to allow students to write the priority and estimated size of the user story (see Figure 2). During the activity, the LEGO constructions should be built by the developers (and only by the developers) using LEGO pieces according to their corresponding user stories and the LEGO construction manuals (see Figure 3). These manuals can also be used by the developers in order to estimate the size of the user stories.

The LEGO Serious Play activity consisted of the following phases:

1. Preparation. Students were divided into teams of 4–6 people and each team appointed a Product Owner and a Scrum Master. Then, each team received a starting pack comprised of the following elements: three instruction manuals (one for the Product Owner, one for the Scrum Master, and one for the developers), a LEGO box with pieces and construction manuals, a set of user story cards, and a kit of Planning Poker cards (see Figure 4). After that, each student read the instructions corresponding to his/her role (Product Owner, Scrum Master, or developer).

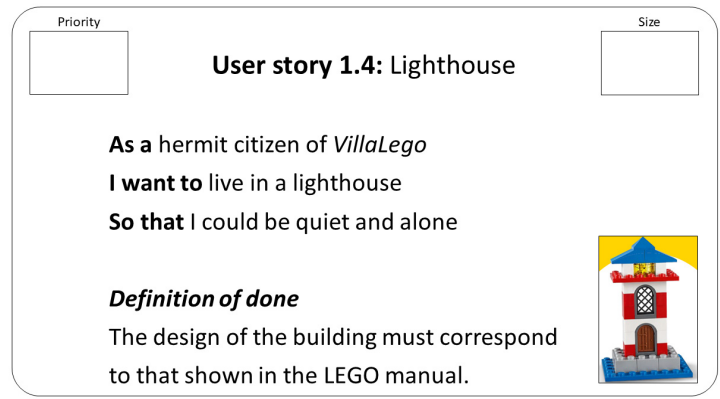

**Figure 2.** Example of a user story card.

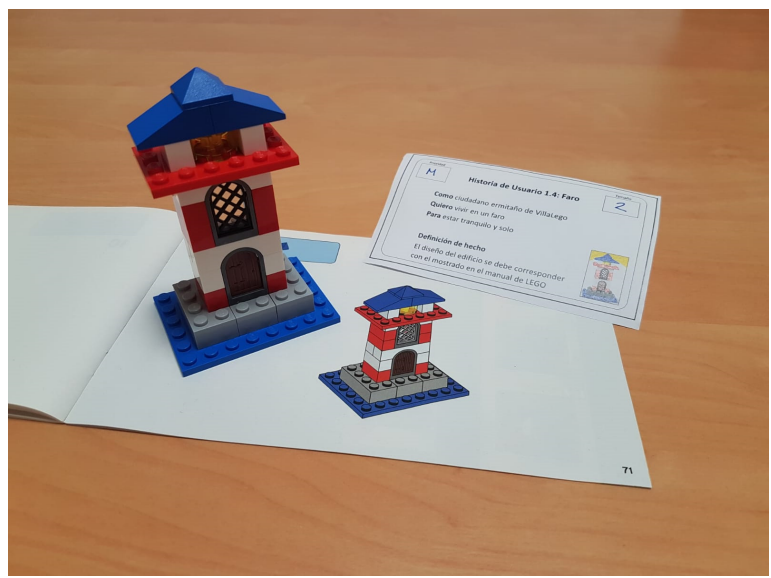

**Figure 3.** Example of a LEGO construction.

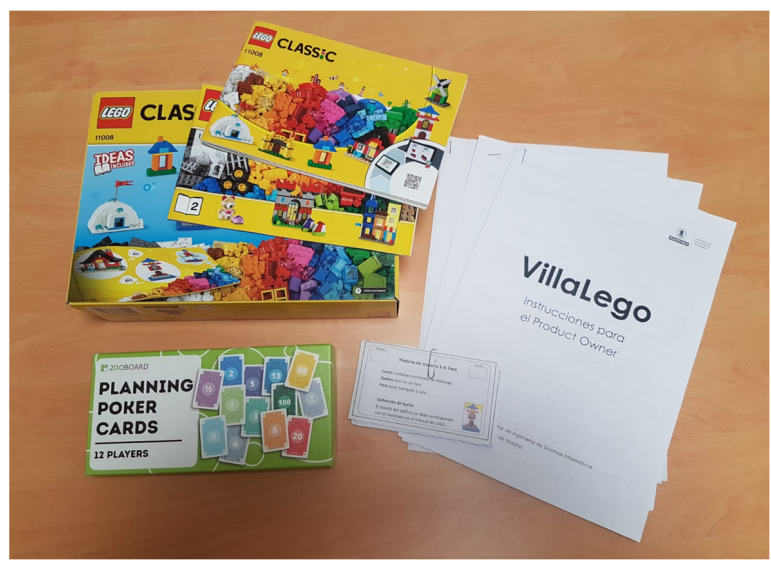

**Figure 4.** LEGO Serious Play activity starting pack.

2. Product backlog prioritization. The Product Owner prioritized the user stories using the MoSCoW technique. The Scrum Master was allowed to help the Product Owner in applying this prioritization technique. Meanwhile, students playing the role of developers continued to familiarize themselves with its instructions.

3.  Sprint planning. The Product Owner explained the client needs to the rest of the Scrum team and then all students defined a goal for the current Sprint. Then, the developers selected user stories to be included in the current Sprint in a consensual manner with the Product Owner. During this process, the size of the selected user stories was estimated by the developers using story points and Planning Poker.

4.  Sprint execution. Once the Sprint planning event was over, students were given a total of 20 min for this phase. Each team had to hold two Daily Scrums during the Sprint: one at the beginning of this phase and another one 10 min later. During the Sprint, the students assuming the role of developers worked on building LEGO constructions in order to complete user stories, whereas the Product Owner was in charge of validating these constructions (i.e., the increments). The Scrum Master was accountable for the Scrum framework being adopted correctly, that is, as defined in the Scrum Guide.

5.  Sprint review. The students held a Sprint review at the end of the Sprint, in which they presented the results of the Sprint (i.e., the LEGO constructions that were completely built during it) to the teachers, who acted as clients in this event. Furthermore, students discussed about what to do on the next Sprints.

6.  Sprint Retrospective. After the Sprint review, students held a Sprint retrospective, in which they discussed what went well during the Sprint, what problems arose, and how these problems were or were not solved. Moreover, students reflected on their performance during the Sprint by using a burndown chart and identified at least one improvement for the next Sprint.

### *3.6. Data Analysis*

The learning performance was determined as the difference between post-test and pre-test scores. A Shapiro–Wilk test of normality determined that not all collected data were normally distributed and hence non-parametric methods were employed. Within each group, the Wilcoxon signed-ranks test for paired samples was utilized to compare the post-test and pre-test scores, while the Mann–Whitney U test was employed to conduct comparisons between groups. In all comparisons, the correlation coefficient r was employed as the measure of effect size. Following Cohen's guidelines [56], an r value between 0.1 and 0.3 indicates a small effect size, between 0.3 and 0.5 it indicates a medium effect size, while an r value of 0.5 or greater signifies a large effect size. Finally, the mean (M) and the standard deviation (SD) were employed to analyze the results of the perceptions questionnaire. Moreover, the Spearman correlation analysis was performed to explore the relationships among the items of the perceptions questionnaire. The data collected and analyzed in this study are provided in the Supplementary Materials.

## 4. Results

### *4.1. Learning Performance*

Table 2 shows the pre- and post-test scores achieved by the students in group A (who played the virtual reality game) and group B (who played the LEGO Serious Play activity). The learning performance, determined by the difference between post-test scores and pre-test scores, exhibited statistical significance in both groups. Its effect size was large (r = 0.59) in group A and medium (r = 0.35) in group B. These figures indicate that the two game-based learning approaches were effective in terms of knowledge acquisition.

When pre-test scores between groups were compared, a statistically significant difference with a medium effect size (*p*-value = 0.01; r = 0.35) was found in favor of group B. On the contrary, the difference between post-test scores obtained by both groups was non-statistically significant and had a less than small effect size (*p*-value = 0.50; r = 0.09). These results indicate that students in group A had less prior knowledge on Scrum and agile practices than their counterparts in group B but that, after the intervention, all students had similar knowledge on this matter regardless of their group. The comparison of learning performance between groups shows that there is a statistically significant difference with a medium effect size (*p*-value = 0.02, r = 0.30) in favor of group A. In view of this fact, it can be concluded that the serious game based on virtual reality used by the students in group A was more effective in terms of knowledge acquisition than the activity based on LEGO Serious Play conducted by the students in group B.

**Table 2.** Pre- and post-test scores.

| Group | Pre-Test | | Post-Test | | Learning Performance | | Wilcoxon Signed-Ranks Test for Paired Samples | |
|---|---|---|---|---|---|---|---|---|
| | **M** | **SD** | **M** | **SD** | **M** | **SD** | ***p*-Value** | **Effect Size (r)** |
| A (N = 22) | 3.9 | 1.8 | 7.3 | 1.3 | 3.5 | 2.3 | <0.01 | 0.59 |
| B (N = 37) | 5.3 | 2.5 | 6.8 | 2.2 | 1.5 | 2.6 | <0.01 | 0.35 |

*4.2. Students Perceptions*

On the one hand, Table 3 shows the Likert items of the questionnaire that were used to collect the students' perceptions toward the game-based learning interventions. On the other hand, Table 4 shows the results of the questionnaire for both groups, including the rating difference between groups for each item.

**Table 3.** Likert items of the questionnaire.

| Item | |
|---|---|
| 1 | My overall opinion of the activity is positive. |
| 2 | The activity helped me learn. |
| 3 | The activity was appealing and motivating. |
| 4 | The activity made learning fun. |
| 5 | The activity was immersive. |
| 6 | The activity was easy to complete. |
| 7 | I needed help to complete the activity. |
| 8 | The activity was well organized. |
| 9 | The activity is useful to learn about Scrum. |
| 10 | The activity is useful to learn about the roles defined by Scrum. |
| 11 | The activity is useful to learn about the events defined by Scrum. |
| 12 | The activity is useful to learn about the artifacts defined by Scrum. |
| 13 | The activity is useful to learn about the Planning Poker technique. |
| 14 | The activity is useful for learning about the MoSCoW technique. |
| 15 | In the future, I would like to conduct activities similar to the one I have conducted in other courses. |
| 16 | I prefer learning through activities such as the one I conducted for learning through traditional teaching methodologies. |
| 17 | The activity is a good complement to traditional teaching methodologies. |
| 18 * | I did not experience any dizziness during the activity. |

* Item 18 was included only for group A.

Overall, the students in both groups had a very positive opinion of the game-based learning activity that was performed in its group and found it beneficial for their learning, as well as motivating, fun, immersive, and well organized. Moreover, in both groups, most students thought the activity was easy to complete and stated that they did not need help to do so. Regarding self-perceived learning effectiveness, in both groups a vast majority of students considered the activity to be useful for learning about Scrum in general and, specifically, for learning about the roles, events, and artifacts defined by Scrum, as well as about two agile practices often used by Scrum teams: MoSCoW, and Planning Poker. Finally, students in both groups were generally in favor of using similar activities in future courses, either as a substitute for traditional teaching or as a complement to it.

**Table 4.** Results of the questionnaire.

| Item | A Group | | B Group | | Mann–Whitney U Test | |
|---|---|---|---|---|---|---|
| | **M** | **SD** | **M** | **SD** | *p*-Value | Effect Size (r) |
| 1 | 4.6 | 0.8 | 4.8 | 0.6 | 0.42 | 0.10 |
| 2 | 4.6 | 0.6 | 4.6 | 0.7 | 0.87 | 0.02 |
| 3 | 4.4 | 0.9 | 4.7 | 0.5 | 0.25 | 0.15 |
| 4 | 4.5 | 0.8 | 4.7 | 0.7 | 0.27 | 0.14 |
| 5 | 4.4 | 0.9 | 4.7 | 0.6 | 0.20 | 0.17 |
| 6 | 4.7 | 0.5 | 4.4 | 0.7 | 0.07 | 0.24 |
| 7 | 1.5 | 1.2 | 2.2 | 1.1 | <0.01 | 0.45 |
| 8 | 4.8 | 0.4 | 4.5 | 0.6 | 0.07 | 0.23 |
| 9 | 4.7 | 0.5 | 4.6 | 0.7 | 0.34 | 0.12 |
| 10 | 4.5 | 0.8 | 4.6 | 0.5 | 0.74 | 0.04 |
| 11 | 4.8 | 0.4 | 4.6 | 0.6 | 0.47 | 0.09 |
| 12 | 4.4 | 0.8 | 4.4 | 0.8 | 0.87 | 0.02 |
| 13 | 4.9 | 0.3 | 4.7 | 0.6 | 0.30 | 0.13 |
| 14 | 4.7 | 0.4 | 4.3 | 1.0 | 0.15 | 0.19 |
| 15 | 4.7 | 0.6 | 4.7 | 0.6 | 0.97 | 0.00 |
| 16 | 3.8 | 1.3 | 4.4 | 1.1 | 0.09 | 0.22 |
| 17 | 4.7 | 0.5 | 4.8 | 0.5 | 0.54 | 0.08 |
| 18 | 3.5 | 1.5 | - | - | - | - |

With respect to the question about dizziness, included only in the questionnaire administered to the students in group A, 7 out of 22 (32%) students reported experiencing some kind of dizziness while using the virtual reality serious game. In this regard, it should be indicated that all participating students were able to complete the activity without major issues. Furthermore, the Spearman correlation analysis was run and no statistically significant correlations were found between dizziness and the intention to use similar virtual reality games in the future or any other item of the perceptions questionnaire.

The average ratings given by the students in both groups were very similar for most questionnaire items. In fact, the effect size of the difference between group ratings was found to be less than small and non-statistically significant at the 0.05 level for all items except for items 6, 7, 8, and 16. For the items 6, 8, and 16 the effect size was small and non-statistically significant. The item 7 (needed help) was the only one for which the difference was found to be statistically significant at the 0.05 level. The effect size of this difference was found to be medium to large (r = 0.45) and indicates that, although in general terms students in both groups agreed that they did not need help to complete the activity, the students in group A agreed more strongly on this statement in comparison with their counterparts. In general, the comments provided by the participating students through the administered questionnaire were in line with the outcomes of the Likert items in both groups. Several students expressed in their comments the innovative nature of both activities and expressed gratitude to the course instructors for them. In group A, some students explained that they felt some dizziness and eyestrain during the use of the virtual reality serious game. In group B, students suggested to increase the time of the activity and to incorporate more Sprints.

## 5. Discussion

The results of this article show that the two game-based learning interventions examined were effective in terms of both knowledge acquisition and motivation. These results are consistent with the current body of research on game-based learning [1–7], which suggests that this methodology is capable of producing positive impacts on student performance and motivation. More specifically, the results obtained for the virtual reality serious game are consistent with those of previous assessments of this game [19–21]. Regarding the results obtained for the LEGO Serious Play activity, it is worth remarking that they are aligned with those of [37–40], who also examined the use of a LEGO Serious Play activity for learning Scrum and found that it was beneficial for the students' learning.

Although student perceptions were very similar for the two interventions, students in group A (who played the virtual reality serious game) outperformed their counterparts in group B (who conducted the LEGO Serious Play activity) in terms of knowledge acquisition. In this regard, it is worth pointing out that the students in group A had less prior knowledge on Scrum and agile practices than those in group B. A possible explanation for this fact is that students in group B had played the virtual reality serious game in the previous academic year in a different course so they received extra training on Scrum and agile practices compared to their peers. The results also show that, in spite of this difference in prior knowledge on Scrum and related agile practices, students in both groups had similar knowledge on this matter after the intervention.

Regarding students' perceptions, it should be remarked that most students found both game-based learning interventions motivating, fun, immersive, easy to complete, and adequately organized, as well as beneficial for their learning. In regard to this latter finding, it is worth pointing out that the students' reports of self-perceived learning effectiveness were aligned with the results of the pre- and post-tests and clearly evidence that both interventions were useful for learning about the core concepts of Scrum including its roles, events, and artifacts, as well as for learning about agile practices frequently used by Scrum teams such as MoSCoW and Planning Poker.

Another interesting finding of this article is that the participating students were generally in favor of conducting similar game-based learning activities in future courses, either as a complement to traditional teaching or as a substitute of traditional teaching. In this regard, it should be noted that the use of such activities as a complement to traditional learning methodologies seems to have greater acceptance among students using them as a replacement. The students who conducted the LEGO Serious Play activity agreed more strongly on these statements than their counterparts, which is a noteworthy but not statistically significant difference between groups. A reasonable explanation for this fact could have been that a significant percentage (32%) of the students who played the virtual reality serious game reported experiencing some kind of dizziness during the intervention. However, all participating students were able to complete the activity without major issues and no statistically significant correlation was found between dizziness and the intention to use similar virtual reality games in the future or any other item of the perceptions questionnaire. Therefore, it can be concluded that students' perceptions toward the virtual reality serious game were not affected by dizziness and that dizziness was not severe in any case.

The only statistically significant difference between groups in terms of student perceptions was that the students who played the virtual reality serious game reported that they did not need help to a greater extent than their counterparts. This is an expected finding, since the serious game is a resource that was designed for self-learning whereas the LEGO Serious Play activity was designed to be conducted in class under the supervision of the teaching staff. Notwithstanding, according to the results of the student questionnaire, most students who participated in the LEGO Serious Play activity did not need help. This finding suggests that the provided instructions, although having room for improvement, were found suitable in most cases.

## 6. Conclusions

This article empirically examined and compared the effectiveness for learning Scrum and related agile practices of a serious game based on virtual reality and a learning activity based on the LEGO Serious Play methodology by means of a quasi-experiment with two groups, pre- and post-tests, and a perceptions questionnaire. The reported results show that both game-based learning approaches were effective in terms of knowledge acquisition and motivation, as well as showing that the students who played the virtual reality serious game outperformed their peers, who conducted a LEGO Serious Play activity, in terms of knowledge acquisition. Therefore, the results of the article suggest that game-based learning using virtual reality serious games or LEGO Serious Play is a suitable option for educators who are willing to teach in an innovative and playful way agile methodologies such as Scrum.

Although previous works examined the use of virtual reality serious games for learning Scrum [19–24], as well as the use of LEGO Serious Play for this same purpose [35,37–42], to the knowledge of the authors this is the first work that performed a comparison between these two game-based learning approaches. Therefore, this article makes a novel contribution by providing, for the first time, evidence on the learning effectiveness of virtual reality serious games compared to LEGO Serious Play. Furthermore, the results reported in this article contribute to a better understanding of the benefits and drawbacks of these two game-based learning approaches.

It should be taken into account that, in spite of providing solid evidence of the effectiveness of virtual reality serious games and LEGO Serious Play, the study presented in this article has some important limitations. First, no random sampling was used because students had to be divided based on whether or not they had played the virtual reality serious game in the past. Second, the study was focused on the evaluation of a single virtual reality serious game and a single LEGO Serious Play activity and the sample size was small (22 for group A and 37 for group B), so the conclusions should be treated with caution. Third, the study examined the learning effectiveness in the short term, but not in the long term. Another noteworthy limitation is that certain aspects of the game-based learning experiences such as enjoyment, motivation, and immersion could have been explored in more detail by extending the perceptions questionnaire to cover additional criteria (e.g., [57]) and by using other evaluations instruments such as the GAMEX scale [58], the Game Engagement Questionnaire [59], the Game Experience Questionnaire [60], or the Igroup Presence Questionnaire [61]. Therefore, an interesting direction for future work is to make further comparisons addressing these aspects in more depth.

An interesting finding of this article was that most students preferred to learn through virtual reality serious games or LEGO Serious Play activities rather than through traditional teaching methods. However, no comparison was conducted between these game-based learning approaches and traditional teaching using control and experimental groups. Thus, future works should conduct this type of comparison, preferably through randomized control trials. Other interesting future works could be to examine the long-term impacts of these approaches through longitudinal studies, as well as to examine them with different games, technologies, instructional designs, and knowledge areas.

**Author Contributions:** Conceptualization, A.G., D.L.-F. and J.M.; methodology, A.G. and D.L.-F.; software, D.L.-F. and J.M.; formal analysis, A.G.; investigation, A.G. and D.L.-F.; resources, A.G. and D.L.-F.; writing—original draft preparation, A.G., D.L.-F. and J.M.; writing—review and editing, A.G., D.L.-F. and J.M. All authors have read and agreed to the published version of the manuscript.

**Funding:** This work was supported in part by Universidad Politécnica de Madrid (UPM) through the educational innovation projects under Grant IE22.6104 and Grant IE23.6104.

**Institutional Review Board Statement:** Ethical review and approval were waived for this study due to the fact that the study only involved the collection of information via surveys and educational tests and all information collected was anonymized so that the identity of the participants cannot be ascertained.

**Informed Consent Statement:** Informed consent was obtained from all subjects involved in the study.

**Data Availability Statement:** The data collected and analyzed in this study are available as Supplementary Material. The authors will request after publication that these data be published under an open license in e-cienciaDatos—https://edatos.consorciomadrono.es (accessed on 17 November 2023).

**Conflicts of Interest:** The authors declare no conflict of interest.

## References

1. Udeozor, C.; Toyoda, R.; Russo Abegão, F.; Glassey, J. Digital games in engineering education: Systematic review and future trends. *Eur. J. Eng. Educ.* **2022**, *48*, 321–339. [CrossRef]
2. Karakoç, B.; Eryılmaz, K.; Turan Özpolat, E.; Yıldırım, İ. The effect of game-based learning on student achievement: A meta-analysis study. *Technol. Knowl. Learn.* **2020**, *27*, 207–222. [CrossRef]
3. Tokac, U.; Novak, E.; Thompson, C.G. Effects of game-based learning on students' mathematics achievement: A meta-analysis. *J. Comput. Assist. Learn.* **2019**, *35*, 407–420. [CrossRef]
4. Hussein, M.H.; Ow, S.H.; Cheong, L.S.; Thong, M.-K.; Ale Ebrahim, N. Effects of digital game-based learning on elementary science learning: A systematic review. *IEEE Access* **2019**, *7*, 62465–62478. [CrossRef]
5. Boyle, E.A.; Hainey, T.; Connolly, T.M.; Gray, G.; Earp, J.; Ott, M.; Lim, T.; Ninaus, M.; Ribeiro, C.; Pereira, J. An update to the systematic literature review of empirical evidence of the impacts and outcomes of computer games and serious games. *Comput. Educ.* **2016**, *94*, 178–192. [CrossRef]
6. Hainey, T.; Connolly, T.M.; Boyle, E.A.; Wilson, A.; Razak, A. A systematic literature review of games-based learning empirical evidence in primary education. *Comput. Educ.* **2016**, *102*, 202–223. [CrossRef]
7. Bodnar, C.A.; Anastasio, D.; Enszer, J.A.; Burkey, D.D. Engineers at play: Games as teaching tools for undergraduate engineering students. *J. Eng. Educ.* **2016**, *105*, 147–200. [CrossRef]
8. Digital.ai. *15th State of Agile Report*; Digital.ai: Raleigh, NC, USA, 2021; Available online: https://info.digital.ai/rs/981-LQX-968/images/SOA15.pdf (accessed on 17 November 2023).
9. Verma, P.; Kumar, R.; Tuteja, J.; Gupta, N. Systematic review of virtual reality & its challenges. In Proceedings of the 3rd International Conference on Intelligent Communication Technologies and Virtual Mobile Networks (ICICV 2021), Tirunelveli, India, 4–6 February 2021; pp. 434–440.
10. Radhakrishnan, U.; Koumaditis, K.; Chinello, F. A systematic review of immersive virtual reality for industrial skills training. *Behav. Inf. Technol.* **2021**, *40*, 1310–1339. [CrossRef]
11. Kurniawan, C.; Rosmansyah, Y.; Dabarsyah, B. A Systematic Literature Review on Virtual Reality for Learning. In Proceedings of the 2019 5th International Conference on Wireless and Telematics, ICWT 2019, Yogyakarta, Indonesia, 25–26 July 2019.
12. Zhang, J.; Chang, J.; Yang, X.; Zhang, J.J. Virtual reality surgery simulation: A survey on patient specific solution. In *Next Generation Computer Animation Techniques, Proceedings of the Third International Workshop, AniNex 2017, Bournemouth, UK, 22–23 June 2017*; Lecture Notes in Computer Science (including subseries Lecture Notes in Artificial Intelligence and Lecture Notes in Bioinformatics); Springer: Cham, Switzerland, 2017; Volume 10582, pp. 220–233.
13. Hernandez, Y.; Ramirez, M.P. Virtual reality systems for training improvement in electrical distribution substations. In Proceedings of the IEEE 16th International Conference on Advanced Learning Technologies, ICALT 2016, Austin, TX, USA, 25–28 July 2016; pp. 75–76.
14. Stănică, I.-C.; Dascalu, M.-I.; Moldoveanu, A.; Bodea, C.-N.; Hostiuc, S. A survey of virtual reality applications as psychotherapeutic tools to treat phobias. In Proceedings of the 12th International Scientific Conference eLearning and Software for Education, Bucharest, Romania, 21–22 April 2016.
15. Cheng, A.; Yang, L.; Andersen, E. Teaching language and culture with a virtual reality game. In Proceedings of the Conference on Human Factors in Computing Systems, Denver, CO, USA, 6–11 May 2017; pp. 541–549.
16. Rivas, D.; Alvarez, M.V.; Guerrero, F.; Grijalva, D.; Loor, S.; Espinoza, J.; Vayas, G.; Huerta, M. Virtual reality applied to physics teaching. In Proceedings of the 2017 9th International Conference on Education Technology and Computers, Barcelona, Spain, 20–22 December 2017; pp. 27–30.
17. Kamińska, D.; Sapiński, T.; Wiak, S.; Tikk, T.; Haamer, R.E.; Avots, E.; Helmi, A.; Ozcinar, C.; Anbarjafari, G. Virtual reality and its applications in education: Survey. *Information* **2019**, *10*, 318. [CrossRef]
18. Pirker, J.; Dengel, A.; Holly, M.; Safikhani, S. Virtual Reality in Computer Science Education: A Systematic Review. In Proceedings of the 26th ACM Symposium on Virtual Reality Software and Technology, Virtual Event, Canada, 1–4 November 2020.
19. Mayor, J.; López-Fernández, D. Scrum VR: Virtual reality serious video game to learn Scrum. *Appl. Sci.* **2021**, *11*, 9015. [CrossRef]
20. López-Fernández, D.; Mayor, J.; Perez, J.; Gordillo, A. Learning and Motivational Impact of Using a Virtual Reality Serious Video Game to Learn Scrum. *IEEE Trans. Games* **2023**, *15*, 430–439. [CrossRef]
21. Lopez-Fernandez, D.; Mayor, J.; Garcia-Perez, M.; Gordillo, A. Are virtual reality serious video games more effective than web video games? *IEEE Comput. Graph. Appl.* **2023**, *43*, 32–42. [CrossRef]
22. Caserman, P.; Gobel, S. Become a Scrum Master: Immersive virtual reality training to learn Scrum framework. In *Serious Games*; Ma, M., Fletcher, B., Göbel, S., Hauge, J.B., Marsh, T., Eds.; Springer: Cham, Switzerland, 2020; pp. 34–48.
23. Radhakrishnan, U.; Koumaditis, K. Teaching Scrum with a Virtual Sprint Simulation: Initial design and considerations. In Proceedings of the 26th ACM Symposium on Virtual Reality Software and Technology, Virtual Event, Canada, 1–4 November 2020; pp. 1–2.

24. Visescu, I.; Blindu, A.; Radhakrishnan, U.; Kadenic, M.; Chinello, F.; Koumaditis, K. Teaching project management in a virtual environment: The Virtual Scrum Simulator (ScrumSim). In Proceedings of the 2022 Nordic Human-Computer Interaction Conference, Aarhus, Denmark, 8–12 October 2022; pp. 1–2.

25. LEGO Serious Play. Available online: https://www.lego.com/themes/serious-play/background (accessed on 17 November 2023).

26. Roos, J.; Victor, B. How It All Began: The Origins Of LEGO® Serious Play®. *Int. J. Manag. Appl. Res.* **2018**, *5*, 326–343. [CrossRef]

27. Dann, S. Facilitating co-creation experience in the classroom with Lego Serious Play. *Australas. Mark. J.* **2018**, *26*, 121–131. [CrossRef]

28. James, A.R. Lego Serious Play: A three-dimensional approach to learning development. *J. Learn. Dev. High. Educ.* **2013**, *6*. [CrossRef]

29. Hansen, P.K.; O'Connor, R. Innovation and learning facilitated by play. In *Encyclopedia of the Sciences of Learning*; Springer: Boston, MA, USA, 2012; pp. 1569–1570.

30. Nielsen, C.B.; Adams, P. Active learning via LEGO MINDSTORMS in Systems Engineering education. In Proceedings of the 2015 IEEE International Symposium on Systems Engineering (ISSE), Rome, Italy, 28–30 September 2015; IEEE: Piscataway, NJ, USA, 2015; pp. 489–495.

31. Kurkovsky, S. Teaching software engineering with LEGO serious play. In Proceedings of the 2015 ACM Conference on Innovation and Technology in Computer Science Education, Vilnius, Lithuania, 4–8 July 2015; pp. 213–218.

32. Kurkovsky, S.; Ludi, S.; Clark, L. Active learning with LEGO for software requirements. In Proceedings of the 50th ACM Technical Symposium on Computer Science Education (SIGCSE 2019), Minneapolis, MN, USA, 27 February 2019; Association for Computing Machinery: New York, NY, USA, 2019; pp. 218–224.

33. Kurkovsky, S. Using LEGO to teach software interfaces and integration. In Proceedings of the 23rd Annual ACM Conference on Innovation and Technology in Computer Science Education, Larnaca, Cyprus, 2–4 July 2018; pp. 371–372.

34. Kurkovsky, S. A LEGO-based approach to introducing test-driven development. In Proceedings of the 2016 ACM Conference on Innovation and Technology in Computer Science Education, Arequipa, Peru, 9–13 July 2016; pp. 246–247.

35. Kurkovsky, S. A simple game to introduce Scrum concepts. In Proceedings of the 51st ACM Technical Symposium on Computer Science Education (SIGCSE 2020), Portland, OR, USA, 11–14 March 2020; p. 1321.

36. López-Fernández, D.; Gordillo, A.; Ortega, F.; Yague, A.; Tovar, E. LEGO® Serious Play in Software Engineering Education. *IEEE Access* **2021**, *9*, 103120–103131. [CrossRef]

37. Krivitsky, A. *lego4scrum: A Complete Guide. A Great Way to Teach the Scrum Framework and Agile Thinking*; Self-published by Alexey Krivitsky: Kyiv, Ukraine, 2017.

38. Velić, M.; Padavić, I.; Dobrović, Ž. Metamodel of Agile Project Management and the Process of Building with LEGO® Bricks. In Proceedings of the 23rd Central European Conference on Information and Intelligent Systems (CECIIS 2012), Varazdin, Croatia, 19–21 September 2012; pp. 481–493.

39. Gama, K. An experience report on using LEGO-based activities in a software engineering course. In Proceedings of the XXXIII Brazilian Symposium on Software Engineering, Salvador, Brazil, 23–27 September 2019; pp. 289–298.

40. Paasivaara, M.; Heikkilä, V.; Lassenius, C.; Toivola, T. Teaching students Scrum using LEGO blocks. In Proceedings of the 36th International Conference on Software Engineering, Hyderabad, India, 31 May–7 June 2014; pp. 382–391.

41. Gama, K.; Oliveira, H. An experience report on teaching Scrum principles in a playful way through distant collaboration with online whiteboards. In Proceedings of the XXXVI Brazilian Symposium on Software Engineering (SBES '22), Virtual, 5–7 October 2022; pp. 143–152.

42. Steghofer, J.-P.; Burden, H. One block on top of the other: Using Minetest to teach Scrum. In Proceedings of the 2022 IEEE/ACM 44th International Conference on Software Engineering: Software Engineering Education and Training, Pittsburgh, PA, USA, 25–27 May 2022; pp. 176–186.

43. Lee, W.L. SCRUM-X: An interactive and experiential learning platform for teaching Scrum. In Proceedings of the 7th International Conference on Education, Training and Informatics (ICETI 2016), Orlando, FL, USA, 8–11 March 2016; pp. 192–197.

44. Christensen, E.L.; Paasivaara, M. Respond to change or die: An educational Scrum simulation for distributed teams. In Proceedings of the ACM/IEEE 44th International Conference on Software Engineering: Software Engineering Education and Training, Pittsburgh, PA, USA, 25–27 May 2022; pp. 235–246.

45. Rodriguez, G.; Soria, Á.; Campo, M. Virtual Scrum: A teaching aid to introduce undergraduate software engineering students to Scrum. *Comput. Appl. Eng. Educ.* **2015**, *23*, 147–156. [CrossRef]

46. Fernandes, J.M.; Sousa, S.M. PlayScrum—A card game to learn the Scrum agile method. In Proceedings of the 2010 Second International Conference on Games and Virtual Worlds for Serious Applications, Braga, Portugal, 25–26 March 2010; pp. 52–59.

47. Marshburn, D.G.; Sieck, J.P. Don't break the build: Developing a Scrum retrospective game. In Proceedings of the 52nd Hawaii International Conference on System Sciences, Maui, HI, USA, 8–11 January 2019; pp. 6988–6996.

48. Von Wangenheim, C.G.; Savi, R.; Borgatto, A.F. SCRUMIA—An educational game for teaching Scrum in computing courses. *J. Syst. Softw.* **2013**, *86*, 2675–2687. [CrossRef]

49. Villavicencio, M.; Narvaez, E.; Izquierdo, E.; Pincay, J. Learning Scrum by doing real-life projects. In Proceedings of the 2017 IEEE Global Engineering Education Conference (EDUCON 2017), Athens, Greece, 25–28 April 2017; pp. 1450–1456.

50. May, J.; York, J.; Lending, D. Play Ball: Bringing Scrum into the classroom. *J. Inf. Syst. Educ.* **2016**, *27*, 87–92.

51.  Schwaber, K.; Sutherland, J. Scrum Guide 2020. Available online: https://scrumguides.org/docs/scrumguide/v2020/2020-Scrum-Guide-US.pdf (accessed on 17 November 2023).
52.  Agile Alliance Agile Alliance Web Portal. Available online: https://www.agilealliance.org (accessed on 17 November 2023).
53.  Bishop, I.; Rizwan Abid, M. Survey of locomotion systems in virtual reality. In Proceedings of the 2nd International Conference on Information System and Data Mining, Lakeland, FL, USA, 9–11 April 2018; pp. 151–154.
54.  Dumlu, B.N.; Demir, Y. Analyzing the user experience of virtual reality storytelling with visual and aural stimuli. In *Design, User Experience, and Usability. Design for Contemporary Interactive Environments, Proceedings of the 22nd HCI International Conference, HCII 2020, Copenhagen, Denmark, 19–24 July 2020*; Lecture Notes in Computer Science; Marcus, A., Rosenzweig, E., Eds.; Springer: Cham, Switzerland, 2020; Volume 12201.
55.  Rangaswamy, S. Visual storytelling through lighting. In Proceedings of the Game Developers Conference, San Jose, CA, USA, 8–12 March 2000.
56.  Cohen, J. *Statistical Power Analysis for the Behavioral Sciences*, 2nd ed.; Routledge: New York, NY, USA, 1988.
57.  Caserman, P.; Hoffmann, K.; Müller, P.; Schaub, M.; Straßburg, K.; Wiemeyer, J.; Bruder, R.; Göbel, S. Quality criteria for serious games: Serious part, game part, and balance. *JMIR Serious Games* **2020**, *8*, e19037. [CrossRef]
58.  Eppmann, R.; Bekk, M.; Klein, K. Gameful experience in gamification: Construction and validation of a gameful experience scale [GAMEX]. *J. Interact. Mark.* **2018**, *43*, 98–115. [CrossRef]
59.  Brockmyer, J.H.; Fox, C.M.; Curtiss, K.A.; McBroom, E.; Burkhart, K.M.; Pidruzny, J.N. The development of the Game Engagement Questionnaire: A measure of engagement in video game-playing. *J. Exp. Soc. Psychol.* **2009**, *45*, 624–634. [CrossRef]
60.  IJsselsteijn, W.A.; de Kort, Y.A.W.; Poels, K. *The Game Experience Questionnaire*; Technische Universiteit Eindhoven: Eindhoven, The Netherlands, 2013.
61.  igroup.org—Project Consortium Igroup Presence Questionnaire (IPQ) Overview. Available online: https://www.igroup.org/pq/ipq (accessed on 17 November 2023).